\documentclass[runningheads]{llncs}

\usepackage[utf8]{inputenc}
\usepackage[T1]{fontenc}
\usepackage{subcaption}
\usepackage{color}
\usepackage{amsmath}
\usepackage{amsfonts}
\usepackage{graphicx}
\usepackage{placeins}
\usepackage{url}
\usepackage{hyperref}
\usepackage[shortcuts]{extdash}
\usepackage{todonotes}

\begin{document}

\title{On the Difficulty of Designing Processor Arrays for Deep Neural Networks}

\author{Kevin Stehle, G\"unther Schindler, Holger Fr\"oning}

\authorrunning{K. Stehle, G. Schindler, H. Fr\"oning}

\institute{Computing Systems Group, Institute of Computer Engineering, Heidelberg University, Germany}

\maketitle

\begin{abstract}
Systolic arrays are a promising computing concept which is in particular inline with CMOS technology trends and linear algebra operations found in the processing of artificial neural networks.
The recent success of such deep learning methods in a wide set of applications has led to a variety of models, which albeit conceptual similar as based on convolutions and fully-connected layers, in detail show a huge diversity in operations due to a large design space:
An operand's dimension varies substantially since it depends on design principles such as receptive field size, number of features, striding, dilating and grouping of features.
Last, recent networks extent previously plain feedforward models by various connectivity, such as in ResNet or DenseNet.
The problem of choosing an optimal systolic array configuration cannot be solved analytically, thus instead methods and tools are required that facilitate a fast and accurate reasoning about optimality in terms of total cycles, utilization, and amount of data movements. 
In this work we introduce \textsc{Camuy}, a lightweight model of a weight-stationary systolic array for linear algebra operations that allows quick explorations of different configurations, such as systolic array dimensions and input/output bitwidths.
\textsc{Camuy} aids accelerator designers in either finding optimal configurations for a particular network architecture or for robust performance across a variety of network architectures.
It offers simple integration into existing machine learning tool stacks (e.g TensorFlow) through custom operators.
We present an analysis of popular DNN models to illustrate how it can estimate required cycles, data movement costs, as well as systolic array utilization, and show how the progress in network architecture design impacts the efficiency of inference on accelerators based on systolic arrays.

\end{abstract}


\section{Introduction}


Deep learning techniques are continuously being applied to more and more applications, with different tasks such as image, signal or speech processing, and deployments including resource-constrained environments such as edge or mobile computing.
As a result, one can observe an increase in diversity in network architectures, which addresses application-specific requirements as well as operational constraints.
This diversity is substantially different to commonly known plain feedforward convolutional networks, including in particular specialized convolutions, such as dilated, grouped or separable ones, extended by various connectivity (ResNet, DenseNet), and vectorized operators (Capsule Networks).
While such network architectures continue to demand for more processing performance, due to technical constraints CMOS scaling is staggering in terms of feature size and absolute performance, among others.
A direct result of this trend is massive parallelization, with systolic arrays being a promising candidate.
A systolic arrays is a regular linear or planar array of processing elements (PE), where each PE may only exchange data and instructions with neighboring PEs. 
Contrary to a pipeline stage in common processors, a PE has more autonomy including local storage and possibly some limited independent control flow.
Systolic arrays are in particular promising as they are a simple abstraction of a massive amount of parallelism, minimizing orchestration overhead in spite of thousands of instructions executing in parallel.

Still, an optimal match between application and processor architecture requires specialization, with particular attention on array size and memory provisioning.
As the reasoning about such architectural parameters is neither intuitive nor trivial, there is a demand for methods and tools to assist designers in finding good or even optimal configurations.

Principally, such predictive modeling can be separated into different concepts~\cite{aspen}.
Analytical models, including rather informal back-of-the-envelope calculations but also methodologically sound methods, are fast, easy to use, flexible and scalable but usually inaccurate.
Contrary, simulations are much more accurate but result in slow down of about 5-6 orders of magnitude.
Furthermore, they are limited in scaling, flexibility, and ease of use (considering massive architectural changes as often required in design space explorations).
Simulations can be further distinguished into functional ones~\cite{Gemmini}, and cycle-accurate ones~\cite{samajdar2018scale}, which mainly differ in speed and accuracy.
An emulation are often considered as another option, which differs from simulation with regard to how behavior is replicated.
In simulation, software is being employed for an exact replication of the original system, while in emulation a different system, not necessarily based on software, is used to replicate internal functions and their relations.
A good example for emulation is the acceleration of simulations by relying on reconfigurable hardware to mimic the behavior of the original system, as often being done by implementing a processor architecture on an FPGA (e.g.,~\cite{Gemmini}).
Last, there also exist various hybrid concepts, for instance Aladdin which combines architecture-level core and memory hierarchy simulations with RTL models and even optimization passes.
Still, Aladdin is closer to simulation in terms of flexibility and speed than analytical models or emulation.

In particular for quick explorations of fundamental design parameters, such as array size and/or memory provisioning, one can observe a lack of tools.
As for such a setting we assume speed being of higher importance than accuracy, we consider emulation as a promising concept.
However, to avoid the implications of FPGA prototyping, we rely on standard CPUs as emulation platform, and instead leverage an identical platform to directly connect the emulator to TensorFlow as a representative for a pervasively used machine learning framework.

We introduce an emulation framework called \textsc{Camuy} - \textit{Configurable Accelerator Modeling for (workload) Understanding and AnalYsis}
, which implements computations using (fast) CPU instructions and focuses on reporting abstract performance metrics such as total cycle count, utilization, and data movements.
Thereby, \textsc{Camuy} allows for a quick and sufficiently accurate exploration of design alternatives, and in particular can assess the suitability of a given architecture (network or processor) to its counterpart.
The main contributions of this work are as follows:

\begin{enumerate}
\item 	Design of technology-agnostic emulation concept and TF integration, reporting abstract metrics such as cycle count, data movements, utilization, and others.
\item 	To demonstrate the effectiveness of \textsc{Camuy}, we detail about the implications of various network architectures for different arrays configurations.
\item 	Furthermore, we provide recommendations on configurations which are robust across a variety of network architectures.
\end{enumerate}

While neglecting many architectural details, such as implementation of floating-point units, on-chip network contention, bank conflicts when accessing on-chip memory, \textsc{Camuy} is focused on fast explorations which are later being refined using other tools. 
\textsc{Camuy} is publicly available\footnote{\url{https://github.com/UniHD-CEG/Camuy}}.



\section{Related Work}


Systolic arrays were introduced in the early 1980s~\cite{1653825} and have recently gained increasing interest for deep learning.
There exist a large variety of systolic implementations, ranging from commercial (TPU~\cite{jouppi2017datacenter}) to research variants (Eyeriss~\cite{chen2016eyeriss0}, Gemmini~\cite{Gemmini}).
\textsc{Camuy} is inspired and modeled following the TPU design, however, it is also extendable to other systolic concepts.

A few exploration tools for systolic arrays have been proposed in the context of deep learning:
SCALE-SIM~\cite{samajdar2018scale} is a configurable, systolic-array-based, cycle-accurate simulator for design space exploration, which is based on a never-stalling array for which traces of read and write addresses are generated, which are then parsed to derive execution time.
Gemmini~\cite{Gemmini} is a systolic array generator that generates a custom ASIC design based on RISC-V and user-defined parameterization, such as bit width and array dimensions.
Evaluation is performed using a cycle-accurate FPGA-accelerated simulation platform.
Further exploration tools, but not centric to systolic arrays, include for instance Aladdin~\cite{shao2014aladdin}, which is a power-performance accelerator modeling framework for system-on-chip (SoC) simulations.
Similarly, Cash~\cite{10.1145/3373087.3375340} is a hardware-software co-design framework for rapid SoC prototyping.

\textsc{Camuy} differs from previous efforts as it focuses on quick and easy explorations of legacy or emerging neural architectures through a simple Tensorflow extension without the need for specialized hardware or software.
The tool aims to assist deep learning experts to develop neural architectures that fit well onto a certain processor array and hardware experts to design processor arrays based on certain neural architectures.
Consequently, \textsc{Camuy} adds an important link between high-productivity machine learning frameworks, such as TensorFlow, and hardware evaluation, for instance based on Gemmini or a TPU.


\section{Emulator Design}

This work is primarily concerned with design space explorations, in order to quickly assess the suitability of a given architecture configuration for a particular DNN or DNN mix.
For such an exploration it is sufficient to assess this suitability based on a set of abstract metrics,
thus emulation is chosen as underlying method, which is in particular fast in comparison to simulations.

We furthermore focus this work on systolic arrays, which are a variant of massively parallel processor arrays, very suitable for regular problems such as linear algebra operations, and a promising candidate to address the increasing costs of data movements. 
Systolic arrays minimize instruction fetch costs, which can otherwise be prohibitive given the vast amount of PEs, and the constrained data flow also avoids network contention effects, which otherwise can be substantial given the non-scalable bisection bandwidth of n-dimensional meshes.
However, data movements can only be reduced if locality effects are sufficiently exploited, and the data flow constraints of a systolic array can result in under-utilization and latency increase.
In this regard, metrics of primary interest include latency in cycles, number of data transfers in between PEs and in and out of the systolic array, resulting bandwidth requirements for a stall-free execution, and utilization as a result of operator size, array dimensions and sparsity.

In more detail, in this work we use a weight-stationary dataflow concept, similar to Google's TPUv1~\cite{jouppi2017datacenter}. While this is controversially discussed~\cite{chen2016eyeriss0,OverratedDataflowChoice}, principally \textsc{Camuy} can also support other concepts.
In this regard, while this work is mainly concerned with exploring the implications of different array dimensions, further architecture configuration parameters of interest include other dataflow concepts, PE SIMDification (multiple multiply-accumulate units per PE), local PE memory (varying with dataflow concept), and the sizing of memory structures such as FIFOs, accumulator array, and global memory.

\begin{figure*}[!htbp]
	\centering
	\includegraphics[width=\textwidth]{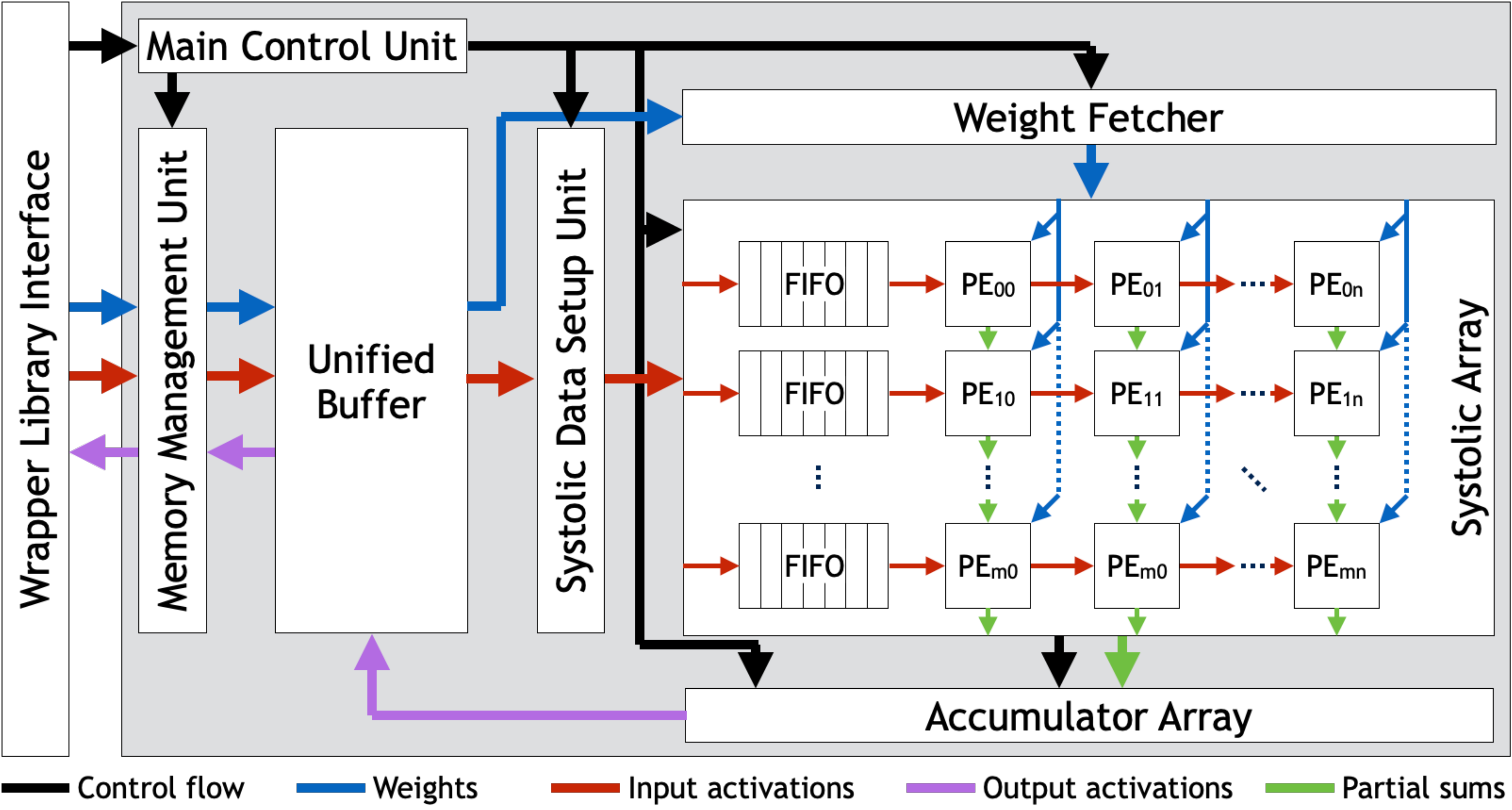}
	\caption{Overview of the CAMUY processor, with systolic array, memory modules, and auxiliary units.}
	\label{fig:camuy-schematic}
\end{figure*}

The basic architecture of \textsc{Camuy} is modeled after the TPUv1~\cite{jouppi2017datacenter} and summarized in Figure~\ref{fig:camuy-schematic}. Notably some alterations were made to better suit the intended use case in resource-constrained environments, such as including only on-chip memory (\textit{Unified Buffer}) for weights, input and output activations.
In the TPUv1, on-chip memory was only used for input and output activations.
The core is a weight-stationary \textit{Systolic Array} of parameterized width $n$ and height $m$.
Note that control flow within the array is not shown for readability reasons.
Its PEs perform MAC operations using partial sums, weights, and activations loaded from neighboring PEs respectively input FIFOs for the boundary PEs.
Thus, each PE only requires 4 data registers: two weight registers to support double buffering, one activation register, and output register for the partial sum.
This register arrangement is a variant of the arrangement proposed by Kung et al.~\cite{kung1980algorithms}.
As for the TPUv1, activations and partial sums flow horizontally respectively vertically through the PE array.
Similar applies to the dataflow from memory to the array, with a \textit{Weight Fetcher} moving weight matrix tiles.
Note hat for a stall-free execution, multiple concurrent weight updates might be necessary, thus multiple such weight updates might be necessary at the same time to ensure stall-free execution, thus
our model allows an arbitrary amount of simultaneous updates and reports this concurrency in terms of bandwidth requirements.
The flow of activations from memory to the PEs is managed by the \textit{Systolic Data Setup Unit}, which fetches one activation row to the FIFOs in a way that waveform requirements are ensured. 
Again following the TPUv1 design, the partial sums produced by the array are accumulated before writing them back to memory (\textit{Accumulator Array}).
While not principally required as neither pipelined activation nor pooling/normalization stages are implemented, it substantially reduces the associate bandwidth requirements.
The \textit{Main Control Unit} orchestrates the different units, in particular for a pipelined and overlapped execution of fetching weight matrix tiles and input activations, performing the systolic operation, and writing back output activations.
Additionally, it controls the \textit{Memory Management Unit}, which transfers data in and out of this processor.

To simplify the integration of the emulator into existing machine learning frameworks, we implemented a wrapper library that dynamically creates emulator instances of certain configurations (bit widths for weights, input and output activations, array dimensions, and accumulator array size).


\section{Evaluation of Network Architectures}

\subsection{Case Study: ResNet-152}
\label{subsec:case-study-resnet-152}

The emulator can guide accelerator developers while searching for optimal systolic array configurations for their respective applications.
In this section, we exemplify the process of finding such optimal configurations using Pareto optimum on the example of a ResNet-152 model with $224\times224$ input images. 
The Pareto optimum is calculated for data movement cost and utilization, both with respect to the total number of cycles required for inference. 
The estimation of data movement cost is performed using the equation

\begin{equation}
	E = 6M_{UB} + 2(M_{INTER\_PE} + M_{AA}) + M_{INTRA\_PE}
	\label{eq:data-movement-energy-cost}
\end{equation}

derived from the estimations performed by Chen et al.~\cite{chen2016eyeriss0}. $M_{UB}$ is the total amount of read and write accesses to the unified buffer. 
The amount of read accesses of PE to registers of neighboring PEs is given by $M_{INTER\_PE}$. $M_{INTRA\_PE}$ is the amount of register read and write accesses inside a PE and $M_{AA}$ is the amount of data movements from the systolic array to the accumulator array. 
The dimensionless normalized total energy movement cost $E$ can be used to compare the energy cost resulting from data movements between CAMUY configurations.

The explored space of systolic array sizes are all possible width and height combinations from 16 to 256 in increments of 8, for a total of 961 possible dimensions.

\begin{figure*}[h!]
	\centering
	\vspace*{-4mm}
	\includegraphics[width=\textwidth]{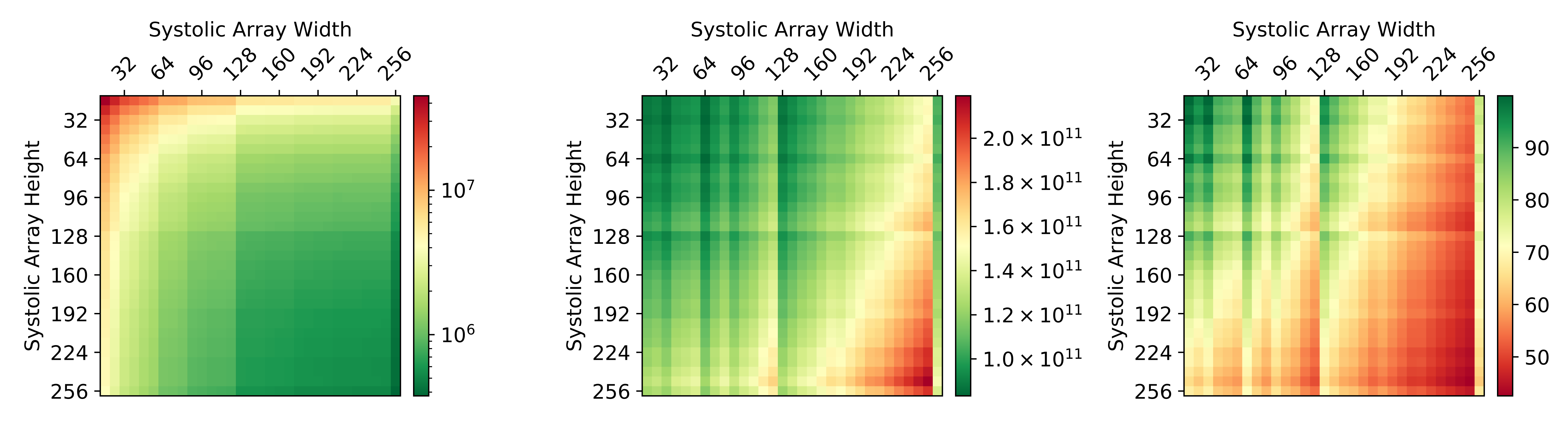}
	\begin{subfigure}{.32\linewidth}
		\caption{Cycle Count}
		\label{subfig:cycle-count-cost-heatmap-resnet-152}
	\end{subfigure}
	\hspace*{0.01\linewidth}
	\begin{subfigure}{.31\linewidth}
		\caption{Data Movement Cost}
		\label{subfig:data-movement-cost-heatmap-resnet-152}
	\end{subfigure}
	\hspace*{0.02\linewidth}
	\begin{subfigure}{.3\linewidth}
		\caption{Utilization}
		\label{subfig:utilization-heatmap-resnet-152}
	\end{subfigure}
	\vspace*{-1mm}
	\caption{Data movement cost and utilization for inference of ResNet-152 using different systolic array dimensions.}
	\vspace*{-4mm}
	\label{fig:heatmaps-resnet-152}
\end{figure*}

Figure~\ref{fig:heatmaps-resnet-152} shows an overview of the measured data movement cost and utilization of the analyzed dimensions in the form of heatmaps in order to give insights on the affects in performance when scaling the two array dimensions.
As can be seen, data movement cost is more sensitive to scaling the array's height than width while the utilization is highly sensitive to the scaling of both.
Furthermore, systolic configurations which are powers of two show a particularly good utilization. 

\begin{figure*}[!htbp]
	\centering
	\includegraphics[width=\textwidth]{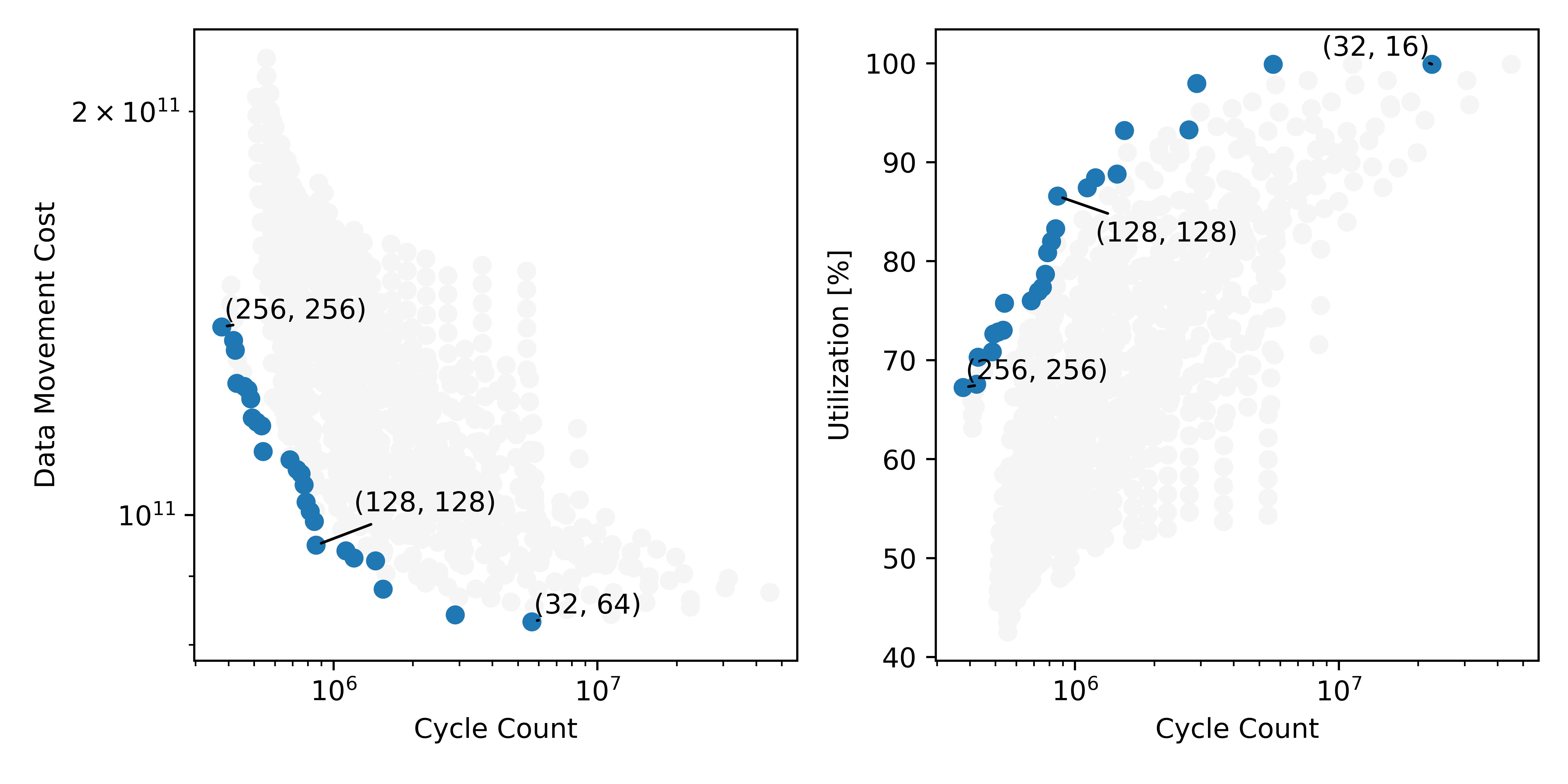}
	\hspace*{.10\linewidth}
	\begin{subfigure}{.44\linewidth}
		\caption{Data Movement Cost}
		\label{subfig:data-movement-cost-to-cycle-count-scatterplot-resnet-152}
	\end{subfigure}
	\begin{subfigure}{.44\linewidth}
		\caption{Utilization}
		\label{subfig:utilization-to-cycle-count-scatterplot-resnet-152}
	\end{subfigure}
	\vspace*{-3mm}
	\caption{Blue dots: Pareto set for data movement cost to cycle count and utilization to cycle count for ResNet-152. Gray dots: Non-optimal dimensions.}
	\vspace*{-4mm}
	\label{fig:scatterplots-combined-resnet-152-small}
\end{figure*}

Next, we calculate the Pareto set using NSGA-II~\cite{deb2002fast}, a multi-objective optimization algorithm, in order to find configurations that are optimal for data movement cost and utilization with respect to the amount of cycle.
Figure~\ref{fig:scatterplots-combined-resnet-152-small} shows the set of all dimensions, with the Pareto frontier highlighted in blue, and the array sizes as annotations in the format \textit{(height, width)}. 
Based on the Pareto sets of the analyzed model, the user is able to choose configurations that best fit their particular use case, be it lowest energy consumption, fastest execution, or a trade-off.

\subsection{Impact of DNN Architecture Development on Systolic Array Performance}
\label{subsec:dnn_architecture}

We conduct further performance analyses (as described in Subsection~\ref{subsec:case-study-resnet-152}) for a wide range of CNN models to show the impact of evolving neural architectures on parameter optimality of systolic array configurations.
We focus on the most popular CNN models as well as their implications on the operand's dimensions for the required matrix multiplication, and show the difficulty of choosing an optimal systolic array configuration.

AlexNet~\cite{krizhevsky2012imagenet} and VGG-16~\cite{simonyan2014very} are classic CNN architectures where the operand's dimension for the required matrix multiplication only depends on the amount of filters and receptive field size.
GoogLeNet and Inception-v2 improved these straight-forward models by applying different receptive fields ($1\times1$, $3\times3$ and $5\times5$) on the same features, thereby increasing variance in the operand's dimension.

A major breakthrough in neural architectures is the introduction of advanced connectivity between layers which leads to significant improvements in parameter and computation efficiency.
The connectivity through residual connections (such as in ResNet~\cite{DBLP:journals/corr/HeZRS15}) enables much deeper models with thinner layers by reducing the vanishing gradient problem, but also results in a reduced operand's dimension for the matrix multiplication.
Dense connectivity (such as in DenseNet~\cite{DBLP:journals/corr/HuangLW16a}) allows even deeper models, where the amount of filters per layer is increased linearly with the model's depth, causing high diversity in the operand's dimensions.

\begin{figure*}[h!]
	\centering
	\vspace*{-6mm}
	\includegraphics[width=\textwidth]{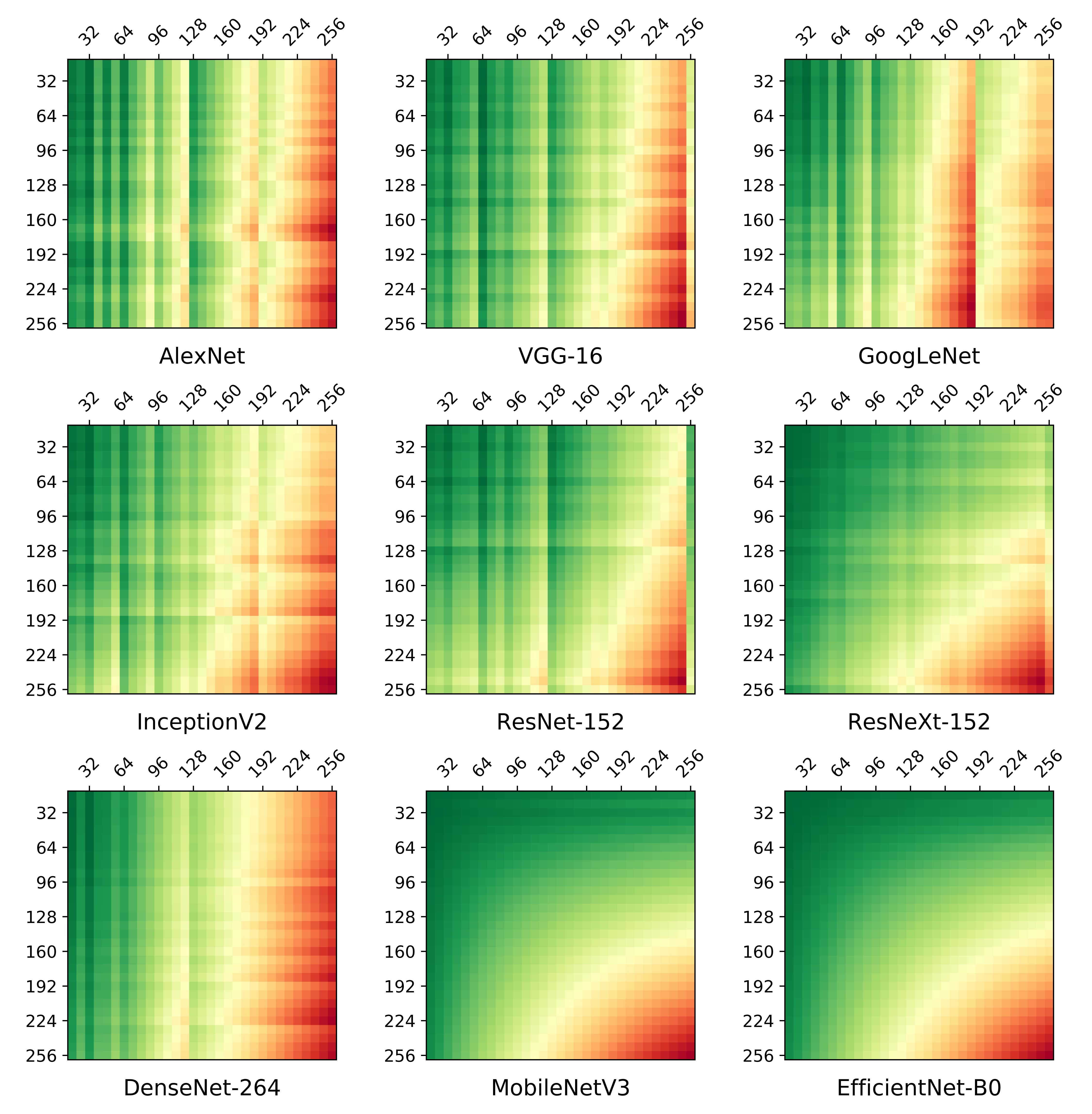}
	\caption{Comparison of data movement heatmaps for different CNN models, with varying dimensions (height on y axis, width on x axis) of the systolic array. 
	Data movement costs are visualized using the spectrum from green (low data movement cost) over yellow (medium data movement cost) to red (high data movement cost).}
	\vspace*{-5mm}
	\label{fig:heatmap-comparison}
\end{figure*}

All state-of-the-art CNN models heavily use group convolutions in order to increase parameter and computation efficiency. 
This grouping, however, leads to a serialization of matrix multiplications (one per group), where the operand's dimensions vary with group size $g$.
We use ResNeXt-152~\cite{xie2017aggregated} with $g=32$ as well as MobileNetV3 and EfficientNet-B0 with $g=1$ (depthwise convolution) as representative for models with group convolution.

We report results from these various CNN models for systolic array configurations of varying height and width, and focus in this work on data movement costs.
This choice is based on the importance of energy efficiency, that data movements are orders of magnitude more energy consuming than arithmetic operations~\cite{horowitz20141,chen2016eyeriss0}, and that systolic arrays are in particular suitable to minimize data movements.


As can be seen in Figure~\ref{fig:heatmap-comparison}, all models are more sensitive to increasing the systolic array's width than the height, indicating that an optimal array configuration is not quadratic.
Comparing residual with dense connections, one can observe that residual ones favor larger array sizes while dense connections benefit from smaller arrays.
Furthermore, dense connections seems to be rather unaffected by varying height, while width matters much more.
Contrary, residual connections are equally sensitive to height and width.

Similar applies to models with group convolutions, where smaller arrays are clearly more beneficial.
Last, all but models with group convolution show an advantage if the width is a power of two (at least), while for group convolutions no such effect is visible.
For models such as AlexNet, GoogLeNet, BN-Inception, VGG-16, and ResNet-152, a similar effect on the height is apparent.

The overall observation is that the inference of almost all analyzed CNN models is significantly more efficient for small systolic arrays and especially for arrays with a low width-to-height ratio.
This finding is in contrast to the commercially available TPU, whose systolic array is quadratic with an edge length of $256$, and furthermore conflictive with the need for parallelization as main technique to further reduce processing time.
\FloatBarrier


\section{Robust/Optimal Processor Architecture Configuration}

In Section~\ref{subsec:dnn_architecture}, we analyzed tendencies of several popular CNN models using data movement costs for varying systolic array configurations.
Next, we aim to find configurations that delivers robust performance in terms of data movement costs and required cycles across the variety of models.

\begin{figure*}[!htbp]
	\centering
	\vspace*{-2.5mm}
	\includegraphics[width=0.8\textwidth]{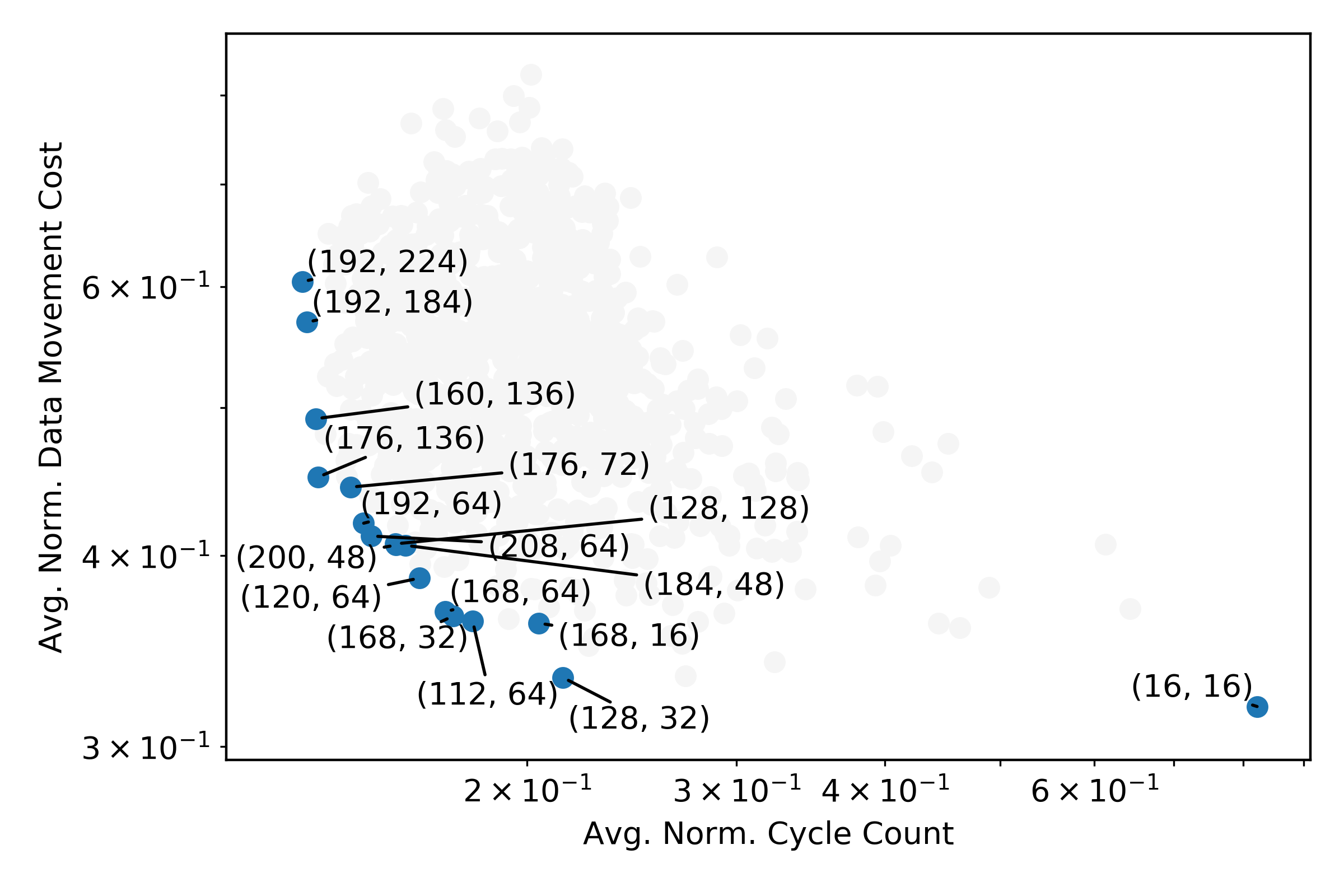}
	\vspace*{-2.5mm}
	\caption{Pareto optimal systolic array dimensions \textit{(height, width)} with regards to average normalized data movement cost relative to latency (Blue Dots). Gray Dots: Dominated dimensions.}
	\vspace*{-3mm}
	\label{fig:scatterplot-data-movement-cost-average}
\end{figure*}

To find such configurations, multi-variate optimization is performed using the averaged normalized results of all analyzed models.
Figure~\ref{fig:scatterplot-data-movement-cost-average} shows the result of this analysis, with the Pareto frontier highlighted in blue and the array sizes as annotations in the format \textit{(height, width)}. 

A surprising result of this analysis is that the configurations with lowest average cycle count are configurations 
with a width that is larger than the height
which goes against the previously observed trend of lower data movement cost for variants with larger height than width.
On the other hand, the configurations with larger height than width achieve much lower data movement costs, while only increasing the cycle count relatively moderately. 
While the configuration with a systolic array width and height of 16 incurs the least data movement cost, it also requires a significantly increased average cycle count relative to the entry with second lowest data movement cost in the Pareto frontier.

Regarding robustness, the configurations in the lower left corner of Figure~\ref{fig:scatterplot-data-movement-cost-average} should be considered, which are all based on a non-square configuration with height larger than width.
Deviations from such configurations lead to significant increases in either data movement cost or cycle count, with only minor improvements in the other dimension.

While this suggests non-square configurations with height larger than width as best solution, we note that this is based on the data transfer model of Equation~\ref{eq:data-movement-energy-cost} and the underlying relation of different types of data movements.
If this relation changes, for instance due to technology scaling, the result optimality of this analysis will also change.
Future work can for instance evaluate the energy data for 14nm, as reported by Dally et al.~\cite{10.1145/3361682}.

\begin{figure*}[!htbp]
	\centering
	\vspace*{-3mm}
	\includegraphics[width=\textwidth]{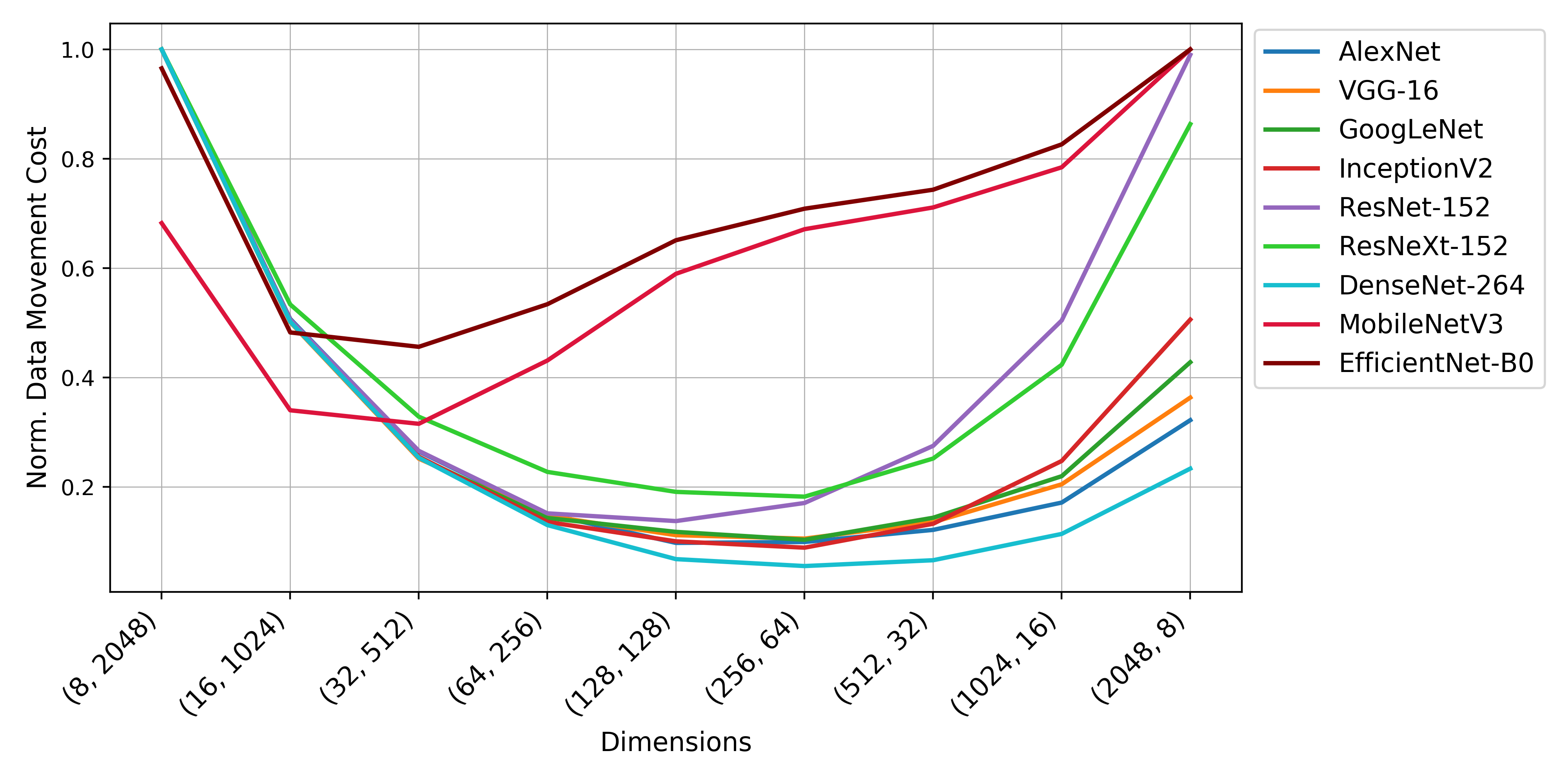}
	\vspace*{-2mm}
	\caption{Normalized data movement cost for inference of analyzed models using systolic array dimensions with equal PE counts.}
	\vspace*{-2mm}
	\label{fig:plot-data-movement-cost-constant-pe-count-combined}
\end{figure*}

The last point of interest is the performance for extreme height to width ratios, in order to evaluate the potential of such configurations. 
For this purpose, the same configuration space is analyzed as done by Samajdar et al.~\cite{samajdar2018scale} in their investigation of weight-stationary systolic array configuration.
The result of this analysis is illustrated in Figure~\ref{fig:plot-data-movement-cost-constant-pe-count-combined} for the average normalized data movement cost.
We find that extreme height to width ratios generally result in low performance, which is inline with similar findings of Samajdar et al. 

\section{Conclusion}
We have presented \textsc{Camuy}, a lightweight model of a weight-stationary systolic array for quick explorations of neural architectures.
The presented tool offers fast and accurate reasoning about optimality in terms of total cycles, utilization, and amount of data movements, in order to reason about the implications of different systolic array configurations, such as optimality or robustness.
For that, we analyzed a large variety of popular CNN models, such as straight-forward, complicated-connected, and grouped architectures.
Our findings suggest that energy and cycle efficiency is optimal for small systolic arrays with a low width-to-height ratio, which is in contrast to recently proposed systolic processors.

Our primary goal for this work is to show the performance gap of modern neural architectures on available systolic hardware, and to further motivate the intersection between hardware design and machine learning.
On that note, we hope that \textsc{Camuy} can assist hardware experts to design processors arrays based on emerging neural designs, and machine learning experts to design neural architectures that fit well onto a certain processor. 

In future work, we will extent \textsc{Camuy} to different systolic concepts, such as output stationary variants and multi-array concepts, in order to improve parallelism for modern CNN models.
Furthermore, we plan to study the impact of emerging and heterogeneous neural architectures, such as transformers and capsule networks, on systolic arrays.

\bibliographystyle{splncs04}
\bibliography{references}

\begin{thebibliography}{10}
\providecommand{\url}[1]{\texttt{#1}}
\providecommand{\urlprefix}{URL }
\providecommand{\doi}[1]{https://doi.org/#1}

\bibitem{chen2016eyeriss0}
Chen, Y.H., Emer, J., Sze, V.: Eyeriss: A spatial architecture for
  energy-efficient dataflow for convolutional neural networks. ACM SIGARCH
  Computer Architecture News  (2016)

\bibitem{10.1145/3361682}
Dally, W.J., Turakhia, Y., Han, S.: Domain-specific hardware accelerators.
  Commun. ACM  (2020)

\bibitem{deb2002fast}
Deb, K., Pratap, A., Agarwal, S., Meyarivan, T.: A fast and elitist
  multiobjective genetic algorithm: {NSGA-II}. IEEE transactions on
  evolutionary computation  (2002)

\bibitem{Gemmini}
{Genc}, H., {Haj-Ali}, A., {Iyer}, V., {Amid}, A., {Mao}, H., {Wright}, J.,
  {Schmidt}, C., {Zhao}, J., {Ou}, A., {Banister}, M., {Shao}, Y.S., {Nikolic},
  B., {Stoica}, I., {Asanovic}, K.: {Gemmini: An Agile Systolic Array Generator
  Enabling Systematic Evaluations of Deep-Learning Architectures}. arXiv
  e-prints  (2019)

\bibitem{DBLP:journals/corr/HeZRS15}
He, K., Zhang, X., Ren, S., Sun, J.: Deep residual learning for image
  recognition. Conference on Computer Vision and Pattern Recognition (CVPR)
  (2016)

\bibitem{horowitz20141}
Horowitz, M.: Computing's energy problem (and what we can do about it). In:
  IEEE International Solid-State Circuits Conference Digest of Technical Papers
  (ISSCC) (2014)

\bibitem{DBLP:journals/corr/HuangLW16a}
Huang, G., Liu, Z., Weinberger, K.Q.: Densely connected convolutional networks.
  Conference on Computer Vision and Pattern Recognition (CVPR)  (2017)

\bibitem{jouppi2017datacenter}
Jouppi, N.P., Young, C., Patil, N., Patterson, D., Agrawal, G., Bajwa, R.,
  Bates, S., Bhatia, S., Boden, N., Borchers, A., et~al.: In-datacenter
  performance analysis of a tensor processing unit. In: ACM/IEEE 44th Annual
  International Symposium on Computer Architecture (ISCA) (2017)

\bibitem{krizhevsky2012imagenet}
Krizhevsky, A., Sutskever, I., Hinton, G.E.: Imagenet classification with deep
  convolutional neural networks. In: Advances in neural information processing
  systems (NIPS) (2012)

\bibitem{1653825}
{Kung}: Why systolic architectures? Computer  (1982)

\bibitem{kung1980algorithms}
Mead, C., Conway, L.: Introduction to VLSI systems. Addison-Wesley (1980)

\bibitem{samajdar2018scale}
Samajdar, A., Zhu, Y., Whatmough, P., Mattina, M., Krishna, T.: Scale-sim:
  Systolic cnn accelerator. arXiv preprint  (2018)

\bibitem{shao2014aladdin}
Shao, Y.S., Reagen, B., Wei, G.Y., Brooks, D.: Aladdin: A pre-rtl,
  power-performance accelerator simulator enabling large design space
  exploration of customized architectures. ACM SIGARCH Computer Architecture
  News  (2014)

\bibitem{simonyan2014very}
Simonyan, K., Zisserman, A.: Very deep convolutional networks for large-scale
  image recognition. International Conference on Learning Representations
  (ICLR)  (2015)

\bibitem{aspen}
{Spafford}, K.L., {Vetter}, J.S.: Aspen: A domain specific language for
  performance modeling. In: Proceedings of the International Conference on High
  Performance Computing, Networking, Storage and Analysis (SC) (2012)

\bibitem{10.1145/3373087.3375340}
Tine, B., Elsabbagh, F., Seyong, L., Vetter, J., Kim, H.: Cash: A single-source
  hardware-software codesign framework for rapid prototyping. In: ACM/SIGDA
  International Symposium on Field-Programmable Gate Arrays (2020)

\bibitem{xie2017aggregated}
Xie, S., Girshick, R., Doll{\'a}r, P., Tu, Z., He, K.: Aggregated residual
  transformations for deep neural networks. In: Conference on computer vision
  and pattern recognition (CVPR) (2017)

\bibitem{OverratedDataflowChoice}
{Yang}, X., {Gao}, M., {Liu}, Q., {Setter}, J.O., {Pu}, J., {Nayak}, A.,
  {Emberton Bell}, S., {Cao}, K., {Ha}, H., {Raina}, P., {Kozyrakis}, C.,
  {Horowitz}, M.: {Interstellar: Using Halide's Scheduling Language to Analyze
  DNN Accelerators}. International Conference on Architectural Support for
  Programming Languages and Operating Systems (ASPLOS)  (2020)

\end{thebibliography}

\end{document}